\documentclass[aps,prl,twocolumn,superscriptaddress,showpacs]{revtex4-1}
\usepackage{graphicx}
\usepackage{amsmath}
\usepackage{amssymb}

\begin{document}
\title{Comment on ``Asymptotic Phase for Stochastic Oscillators''}
\author{Arkady Pikovsky}
\affiliation{Institute for Physics and Astronomy, 
University of Potsdam, Karl-Liebknecht-Str. 24/25, 14476 Potsdam-Golm, Germany}
\affiliation{Department of Control Theory, Nizhni Novgorod State University,
Gagarin Av. 23, 606950, Nizhni Novgorod, Russia}

\begin{abstract}
Definition of the phase of oscillations is straightforward for deterministic 
periodic processes but nontrivial for stochastic ones. Recently, Thomas and Lindner in
[Phys. Rev. Lett., v. 113, 254101 (2014)] suggested  to use the 
argument of the complex eigenfunction
of the backward density evolution operator with the smallest real part of the eigenvalue,
as an asymptotic phase of stochastic oscillations. Here I show that this definition 
does not generally provide a correct asymptotic phase.
\end{abstract}
\date{\today}

\date{\today}

\maketitle

Notion of the phase of periodic oscillators lies at the heart of characterization
of oscillatory processes.
In deterministic systems, the phase on the limit cycle is defined straightforwardly as
a $2\pi$-periodic variable on the cycle 
which grows in time uniformly. For stable oscillations, the
definition can be extended also to the basin of attraction of periodic regime, via
construction of isochrons -- the surfaces of Poincar\'e sections of the flow with
the return time being exactly the period of oscillations. The isochrons are surfaces of 
the constant phase (sometimes called also the asymptotic phase). Recently, in 
\cite{Schwabedal-Pikovsky-13} the notion of isochrons has been extended to stochastic
systems, where they have been defined as the surfaces of constant mean 
first return time.
In paper \cite{Thomas-Lindner-14}, a different definition of the asymptotic phase of 
stochastic oscillations has been suggested, based on the properties of the operator
describing the evolution of the probability density (we assume for simplicity
of presentation that it is the
Fokker-Planck operator). If the nontrivial 
eigenvalue of the backward Fokker-Planck operator with 
least negative real part is complex, the systems has been called robustly 
oscillatory. With some 
further minor technical conditions, for such systems the asymptotic phase  was defined
in \cite{Thomas-Lindner-14} as the argument of the complex eigenfunction corresponding
to the first nontrivial complex eigenvalue. 

In this comment I present a simple analytically solvable example, where the
definition of 
\cite{Thomas-Lindner-14} does not provide a proper phase. I consider a limit cycle
in a three-dimensional phase space, and use poloidal coordinates $(\theta,\phi,r)$ around 
it.  The limit cycle corresponds to $r=0$ and is parametrized by its phase $\theta$, which
is governed by $\dot\theta=\Omega$, where $\Omega$ is the frequency of the oscillations.
Transversal to the limit cycle direction is spanned by variables $(r,\phi)$.
The limit cycle is stable, and the transversal perturbations decay according to
$\dot r=-\delta r$, $\dot\phi=\omega$, where $-\delta$ is the rate of decay 
of perturbations.
In the whole vicinity of the limit cycle, the variable $\theta$ is the proper phase.

We assume now that these oscillations are subject to 
independent Gaussian white noise terms in all variables:
\begin{align*}
\dot\theta&=\Omega+\sigma_\theta\xi_\theta(t)\;,\\
\dot\phi&=\omega+\sigma_\phi\xi_\phi(t)\;,\\
\dot r&=-\delta r+\sigma_r\xi_r(t)\;,
\end{align*}
where $\langle \xi_\theta(t)\xi_\theta(t')\rangle=2\delta(t-t')$
and similarly for other terms. 
If the noise terms are sufficiently small, the oscillations are slightly perturbed
deterministic ones, and are well defined for all relations between the noise 
intensities;
the
approach of Ref.~\cite{Schwabedal-Pikovsky-13} yields here the 
proper phase $\theta=const$.

Let us apply approach of Ref.~\cite{Thomas-Lindner-14}.
An eigenvalue problem for the backward Fokker-Planck
equation
\[
\frac{\partial \rho}{\partial t}=
\Omega\frac{\partial \rho}{\partial\theta}+\omega\frac{\partial\rho}{\partial \phi}-
\delta r\frac{\partial  \rho}{\partial r}
 +\sigma_\theta^2\frac{\partial^2\rho}{\partial\theta^2}
+\sigma_\phi^2\frac{\partial^2\rho}{\partial\phi^2}
+\sigma_r^2\frac{\partial^2\rho}{\partial r^2}
\]
can be solved by separation of variables. The eigenfunctions 
\[
\rho_{nml}=e^{in\theta}e^{im\phi}R_l(r)
\]
are parametrized by three integers $(n,m,l)$, and the eigenvalues are
\[
\lambda_{nml}=in\Omega+im\omega-n\sigma_\theta^2-m\sigma_\phi^2-\beta_l, \qquad n,m,l=0,1,2,\ldots\;,
\]
where  $\beta_l$ are the eigenvalues of the radial equation
\[
\sigma_r^2\frac{d^2R_l}{dr^2}-r\frac{d}{dr}\delta R_l+\beta_lR_l=0\;.
\]
The latter is nothing else but the equation for the Ornstein-Uhlenbeck process,
with the eigenfunctions being Hermite polynomials and the eigenvalues $\beta_l=l\delta$ (see
\cite{Gardiner-96}, section 5.2.6  for details). 

According to \cite{Thomas-Lindner-14}, we focus on the nontrivial eigenvalue with least 
negative  real part. First, we observe that for $\delta$ small compared
to $\sigma_\theta^2,\sigma_\phi^2$, this eigenvalue
is $\lambda_{001}$ and real, so that the process is not robustly oscillatory according
to classification of Ref.~\cite{Thomas-Lindner-14}. If parameter $\delta$ is large, then
this eigenvalue is complex, and one has two possibilities:
\begin{enumerate}
\item If $\sigma_\phi<\sigma_\theta$, then
$
\lambda_{min}=\lambda_{010}=i\omega-\sigma_\phi^2
$.
In this case the eigenfunction is
$
\rho_{010}=R_0(r)e^{i\phi}
$.
\item If $\sigma_\phi>\sigma_\theta$, then
$
\lambda_{min}=\lambda_{100}=i\Omega-\sigma_\theta^2
$.
In this case the eigenfunction is
$
\rho_{100}=R_0(r)e^{i\theta}
$.
\end{enumerate}
One can see that only in case (2) the proper phase $\theta$ is recovered, while 
in case (1) the variable $\phi$, which is transversal to the correct phase $\theta$,
is delivered as an asymptotic phase variable according to the method of 
Ref.~\cite{Thomas-Lindner-14}.  

In conclusion, while the phase is a rotating variable, 
in the phase space of a noisy dynamical
system there can be many rotations corresponding to complex eigenvalues of the density
evolution operator. Therefore one generally cannot identify the phase using the least
stable complex eigenfunction.

\end{document}